\documentclass[aoas,nameyear,dvips]{arximspdf}
\usepackage{multirow}
\usepackage{graphicx}

\doi{10.1214/10-AOAS362}
\volume{4}
\issue{4}
\pubyear{2010}
\firstpage{2099}
\lastpage{2113}

\makeatletter

\newtheorem{theorem}{Theorem}

\renewcommand{\cite}{\citet}
\renewcommand{\upsilon}{v}

\newcommand{\bh}{\hat{\beta}}
\newcommand{\sh}{\hat{\sigma}}
\newcommand{\C}{\operatorname{Cov}}
\newcommand{\D}{\operatorname{diag}}

\newcommand{\argmin}{\operatorname{argmin}}

\makeatother

\begin{document}
\begin{frontmatter}

\title{Model-robust regression and a Bayesian ``sandwich'' estimator}
\runtitle{Model-robust regression and a Bayesian ``sandwich'' estimator}

\begin{aug}
\author[a]{\fnms{Adam A.} \snm{Szpiro}\corref{}\ead[label=e1]{aszpiro@u.washington.edu}},
\author[a]{\fnms{Kenneth M.} \snm{Rice}}
\and
\author[a]{\fnms{Thomas} \snm{Lumley}}
\runauthor{A. A. Szpiro, K. M. Rice and T. Lumley}
\affiliation{University of Washington, University of Washington and University of Washington}
\address[a]{Department of Biostatistics\\
University of Washington\\
Box 357232\\
Seattle, Washington 98195\\
USA} 
\end{aug}

\received{\smonth{4} \syear{2009}}
\revised{\smonth{4} \syear{2010}}

%
\begin{abstract}
We present a new Bayesian approach to model-robust linear regression
that leads to uncertainty estimates with the same robustness properties
as the Huber--White sandwich estimator. The sandwich estimator is known
to provide asymptotically correct frequentist inference, even when
standard modeling assumptions such as linearity and homoscedasticity in
the data-generating mechanism are violated. Our derivation provides a
compelling Bayesian justification for using this simple and popular
tool, and it also clarifies what is being estimated when the
data-generating mechanism is not linear. We demonstrate the
applicability of our approach using a simulation study and health care
cost data from an evaluation of the Washington State Basic Health Plan.
\end{abstract}

%

\end{frontmatter}
%
\section{Introduction}\label{se:intro}

The classical theory of uncorrelated linear regression is based on
three modeling assumptions: (i) the outcome variable is linearly
related to the covariates on average, (ii) random variations from the
linear trend are homoscedastic, and (iii) random variations from the
linear trend are Normally distributed. Under these assumptions,
classical frequentist methods give point estimates and exact
probability statements for the sampling distribution of these
estimates. Equivalent uncertainty estimates are derived in the
Bayesian paradigm, but are stated in terms of the posterior
distribution for the unknown slope parameter in the assumed linear
model. However, in a typical application none of these modeling
assumptions can reasonably be
expected to hold in the data-generating mechanism.

\begin{figure}

\includegraphics{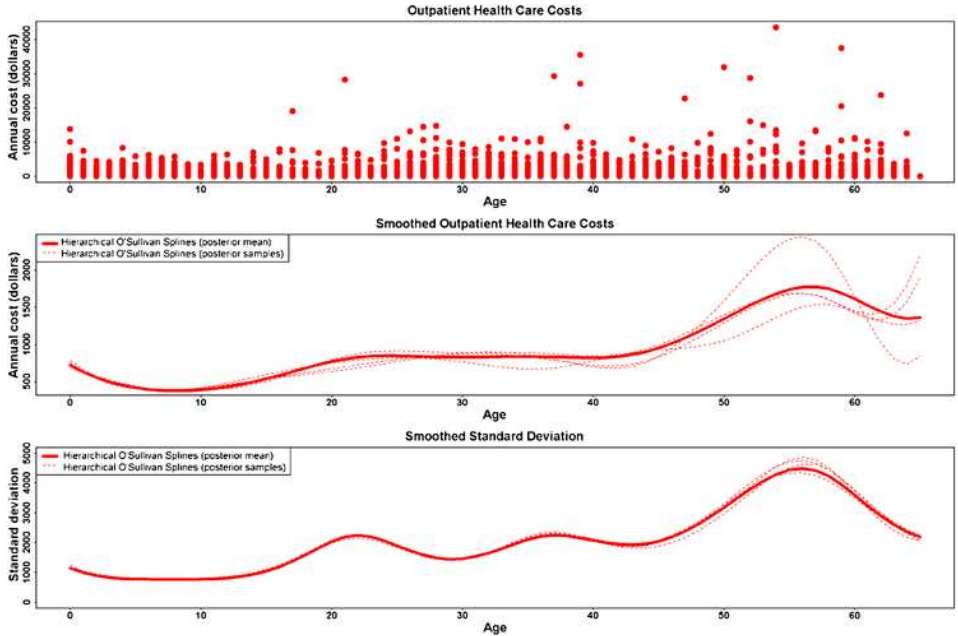}

\caption{Outpatient health care costs from the evaluation of Washington
States Basic Health Plan.  Top panel: Average annual costs for 6918 subjects enrolled in the study.
Middle panel: Semi-parametric smoothing estimates of the average annual cost vs. age, fit with
 Bayesian O'Sullivan splines.  Bottom panel: Semi-parametric smoothing estimate of
the standard deviation of annual health care costs vs. age, fit with Bayesian O'Sullivan splines.
In each of the bottom two panels, the thick red line is the posterior mean of the spline fit, and the thin
dashed red lines are example draws from the posterior distribution.}\label{fi:health}
\end{figure}

\begin{figure}

\includegraphics{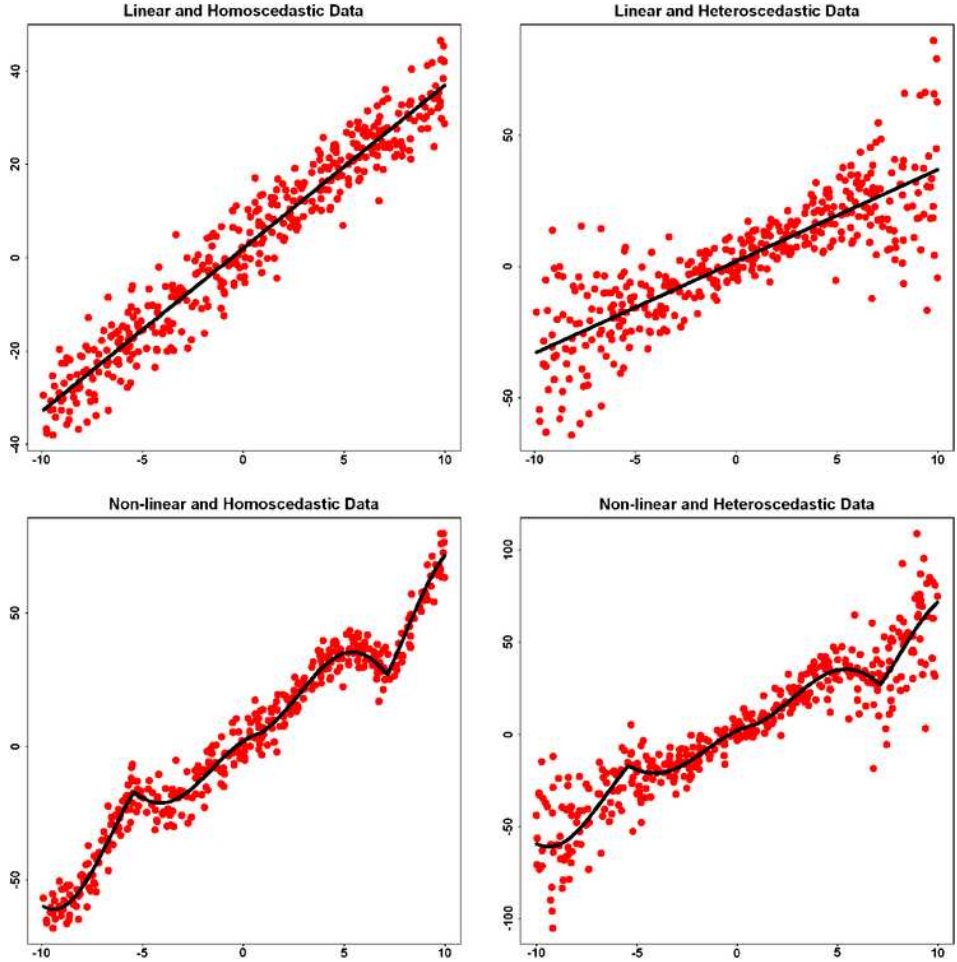}

\caption{Example scatterplots (red dots) and mean functions (black lines)
from the four simulation scenarios
 considered in Section \protect\ref{se:sim} with $n=400$.
The four scatterplots correspond to all possible combination of the
linear and nonlinear mean functions and homoscedastic and
heteroscedastic variance functions defined in Section
\protect\ref{se:sim}.}\label{fi:examples}
\end{figure}

We study the relationship between age and average annual outpatient
health care costs using data from the evaluation of the Washington
State Basic Health Plan. The plan provided subsidized health insurance
for low income residents starting in 1989, and the evaluation study
included 6918 subjects followed for an average of 22 months (range 1 to
44 months) [\cite{Diehr1993}]. Previous analysis of this data set has
shown that the variability is heteroscedastic and not Normally
distributed [\cite{Lumley2002}], and it appears from
Figure \ref{fi:health} that the relationship deviates from linearity.
We are still motivated to estimate a ``linear trend'' since this appears
to be a dominant feature of the data, and while we could consider a
transformation to stabilize the variance, this may not be desirable
since the primary policy interest is in total or mean dollars not in
log-dollars [\cite{Diehr1999}]. We also consider simulated data sets
with similar features as illustrated in Figure~\ref{fi:examples}.

For the classical theory of linear regression to hold, the Normality
assumption is only necessary if we want to derive exact sampling
probabilities for the point estimates. In the large sample limit, the
central limit theorem alone guarantees that the sampling distribution
is asymptotically Normal and that the classical standard error
estimates are correct. The linearity and homoscedasticity assumptions,
however, are a different matter. If either of these is violated in the
data-generating mechanism, then classical standard error estimates are
incorrect, even asymptotically. Furthermore, without the assumption of
linearity, it is not immediately clear what quantity we are trying to
estimate.

A modern frequentist approach to analyzing data that do not conform to
classical assumptions is to directly state what we want to know about
moments of the data-generating mechanism by way of estimating
equations, without making any assumptions about validity of an
underlying model. The associated ``robust'' or ``sandwich''-based standard
errors provide accurate large sample inference, at no more
computational effort than fitting a linear model [\cite{Huber1967};
\cite{White1980}; \cite{Liang1986}; \cite{Royall1986}]. The Huber--White sandwich estimator
is easy to implement with standard software and is widely used in
biostatistics. As long as the data have a dominant linear structure,
this strategy provides relevant inference for the linear trend and does
not depend on detailed modeling of the variance or mean structures.

Finding a Bayesian analogue of estimating equations and the sandwich
estimator has been an open problem for some time. In this paper we
describe a novel Bayesian framework for linear regression that assumes
neither linearity nor homoscedasticity. Even in the absence of a slope
parameter, we give a natural definition for the ``linear trend'' quantity
to be estimated and how to measure its uncertainty. We show that in
the random covariate setting our Bayesian robust posterior standard
deviations are asymptotically equivalent to the commonly used sandwich
estimator. Furthermore, with fixed covariates our Bayesian robust
uncertainty estimates exhibit better frequentist sampling properties
than the sandwich estimator, when the true data-generating mechanism is
nonlinear in the covariates.

In Section \ref{se:notation} we set out our notation and define the
model-robust Bayesian regression paradigm. In
Section \ref{se:discrete} we derive our main theoretical results for
the case of randomly sampled covariates from a discrete space. In
Section \ref{se:extensions} we consider extensions to continuous
covariates and to a fixed design matrix. We demonstrate the properties
of our methodology in a simulation study in Section \ref{se:sim}, and
in Section \ref{se:ex} we apply it to the annual health care cost data
described above. We conclude in Section \ref{se:disc} with a
discussion.

\section{Notation and definitions}\label{se:notation}
\subsection{Target of inference}
We consider the familiar situation for multivariate linear regression
of having observed an $n$-vector of outcomes $Y$ and an $n \times m$
matrix of covariate values $X$, with the stated objective of estimating
the ``linear relationship'' between $X$ and $Y$. Before determining
operationally how to do this, we take care to clarify the quantity of
interest in terms of a true (but unknown) data-generating mechanism,
without assuming that there is an underlying linear relationship.

We assume that $X$ represents $n$ independent identically distributed
observations in $\mathbb{R}^m$ of the $m$-dimensional covariate random variable
$x$, and that $Y$ represents $n$ corresponding independent observations
of the real-valued outcome random variable $y$. We think of the
probability distribution for $x$ as representing the frequency of
different covariate values in the population to which we wish to
generalize, and the distribution of $y$ conditional on $x$ as the
distribution of the outcome for individuals with covariate values $x$.
Suppose that the true joint distribution for $x$ and $y$ admits a
density function $\lambda(\cdot)$ for $x$ (with respect to the Lebesgue
measure on $\mathbb{R}^m$) such that for any measurable set $A$
%
\begin{equation}\label{eq:lambda}
 P(x \in A) = \int_A \lambda(\upsilon)\,d\upsilon
\end{equation}
and a measurable function $\phi(\cdot)$ on $\mathbb{R}^m$ for the mean of $y$
conditional on $x$ such that
%
\begin{equation}\label{eq:phi.star}
E(y|x=\upsilon) =\phi(\upsilon).
\end{equation}
Throughout, we use $\upsilon$ as a dummy variable
for $x$.

Heuristically, we can say that we are interested in the ``linear
relationship'' between $x$ and the true conditional mean of $y$. If
$\phi(\cdot)$ were known to be linear, we would simply be interested in
its coefficients. Since we are not assuming that the true mean function
is linear, one possible approach is to define the quantity of interest
as the $m$-vector of minimizing coefficients from the least-squares
linear fit
%
\begin{equation}\label{eq:beta*}
\beta= \argmin\limits_\alpha\int\bigl(\phi(\upsilon) - \upsilon\alpha\bigr)^2 \lambda(\upsilon)\,d \upsilon.
\end{equation}
We can describe $\beta$ as the set of $m$ coefficients that minimizes
the average squared error over the entire population in approximating
the mean value of $y$ by a linear function of $x$.

The definition of $\beta$ is essentially a statement about the
scientific question of interest, and it is not concerned with the
details of random sampling of the observations. We have identified the
deterministic function $\phi(\cdot)$ as representing the mean
dependence of $y$ on $x$, and our objective is to approximate this
curve by a straight line. We define $\beta$ as the best linear
approximation to the curve $\phi(\cdot)$ by the method of
least-squares, an idea that dates to the early work of
\cite{Gauss1809}, \cite{Legendre1805} and \cite{Jacobi1841}. Our
goal is inference for $\beta$, not for the full function $\phi(\cdot)$.

\cite{Freedman2006} has pointedly described the dangers of fitting a
linear model when such a model does not hold and then deriving ``robust''
standard error estimates for an uninterpretable parameter. Our approach
is fundamentally different in that we explicitly recognize that the
data-generating mechanism may be nonlinear, and we define $\beta$ as a
quantity of interest that summarizes the linear feature in the
data-generating mechanism (this corresponds to the standard definition
of $\beta$ if the data-generating mechanism is linear). While $\beta$
can be defined mathematically in a very general setting, consistent
with the ideas in \cite{Freedman2006}, we recommend it as a relevant
target of inference only when the data suggest a dominant linear trend.

\subsection{Bayesian inference}\label{se:model}

 Since we do not know the true mean function
$\phi(\cdot)$ or the true covariate density $\lambda(\cdot)$, we cannot
directly calculate $\beta$ from equation~(\ref{eq:beta*}), and we need
to take advantage of the observations in order to make inference about
$\beta$. To do this, we embed $\phi(\cdot)$ and $\lambda(\cdot)$
in a
flexible Bayesian model in such a way that we can derive posterior
distributions for these functions and, thus, derive a posterior
distribution for $\beta$. The key consideration in constructing the
Bayesian model is that it be highly flexible, assuming neither
linearity nor homoscedasticity.

We adopt the conditionally Normal model for $y$,
\[
y|x,\phi(\cdot),\sigma^2(\cdot) \sim N(\phi(x), \sigma^2(x)),
\]
where we have introduced the ancillary unknown variance function
$\sigma^2(\cdot)$. To complete the Bayesian model, it remains to
specify a prior distribution, with probability measure $\pi
(\lambda(\cdot), \phi(\cdot), \sigma^2(\cdot))$, which will be chosen
to have a density that can be written
%
\begin{equation}\label{eq:priors}
p(\lambda(\cdot), \phi(\cdot), \sigma^2(\cdot) ) =
p_\lambda(\lambda(\cdot))p_{\phi, \sigma^2} (\phi(\cdot),
\sigma^2(\cdot)).
\end{equation}
We will give specific examples of priors in
the remainder of this paper.

Defining priors for the discrete covariate case is relatively
straightforward because we can specify a saturated model for the mean
and variance functions $\phi(\cdot)$ and $\sigma^2(\cdot)$, and use a
Dirichlet distribution for $\lambda(\cdot)$. We derive our main
theoretical results for that setting in Section \ref{se:discrete}.
Later, in Section \ref{se:extensions}, we present a simple and
effective approach for extending the method to continuous covariates by
using spline-based priors for $\phi(\cdot)$ and $\sigma^2(\cdot)$.

Once we have specified priors in equation (\ref{eq:priors}), standard
Bayesian calculus gives a posterior distribution for $\phi(\cdot)$ and
$\lambda(\cdot)$,
\[
\pi(\lambda(\cdot), \phi(\cdot) |X,Y),
\]
 and therefore a posterior distribution for the $m$-dimensional
vector $\beta$,
%
\begin{equation}\label{eq:beta}
\pi(\beta | X,Y) =
\pi\biggl(\argmin\limits_\alpha\int\bigl(\phi(\upsilon) - \upsilon\alpha\bigr)^2
\lambda(\upsilon)\,d \upsilon \Big| X,Y\biggr).
\end{equation}
Following common
practice, we define a point estimate by taking the posterior mean of
$\beta$,
%
\begin{equation}\label{eq:beta.hat}
\hat{\beta}_j = E_\pi(\beta_j | X,Y),\qquad j=1,\ldots,m,
 \end{equation}
and we use its posterior standard deviation as
a measure of uncertainty
%
\begin{equation}\label{eq:sigma.hat}
\hat{\sigma}_{\beta_j} = \D(\C_\pi(\beta| X,Y) )^{1/2}_j,\qquad j=1,\ldots,m.
 \end{equation}
We can construct approximate moment-based 95\%
credible intervals with the formulation
\[
\mathit{CI}_{{95}_j} =
\hat{\beta}_j \pm1.96 \hat{\sigma}_{\beta_j},\qquad j=1,\ldots,m.
\]
%

\section{Discrete covariates}\label{se:discrete}
In this section we complete the specification of
the Bayesian model for the discrete covariate case and derive our main
theoretical results in that setting. Let $\xi= (\xi_1, \ldots, \xi_K)$
consist of $K$ nonzero deterministic $m$-vectors that span $\mathbb{R}^m$, and
suppose that the covariate $x$ can take these values. Let $n_k$ be the
number of $i=1,\ldots,n$ such that $X_i=\xi_k$, where $X_i$ is the $i$th
row of $X$. We let $\lambda(\cdot)$ be a density with mass restricted
to $\xi\subset\mathbb{R}^m$, written in the form
\[
\lambda(\cdot) =
\sum_{k=1}^K \lambda_k \delta_{\xi_k}(\cdot),
\]
where
$\delta_{\xi_k}$ is the Dirac delta function with point mass at
$\xi_k$. That is,
\[
P\bigl(x=\xi_k;\lambda(\cdot)\bigr) =\lambda_k,\qquad\sum_{k=1}^K\lambda_k = 1.
\]
We use an improper
Dirichlet prior for $\lambda(\cdot)$ such that its density can be
written
\[
p_\lambda(\lambda(\cdot)) \propto\prod_{k=1}^{K}\lambda_k^{-1}\qquad\Biggl(0 \mbox{ if } \sum_{k=1}^K \lambda_k\not= 1\Biggr).
\]
The posterior distribution of $\lambda(\cdot)$ is also
Dirichlet with density
\[
p_{\lambda|X} (\lambda(\cdot))\propto\prod_{k=1}^{K} \lambda_k^{-1+n_k} \qquad\Biggl(0 \mbox{ if }\sum_{k=1}^K \lambda_k \not= 1\Biggr).
\]
One way to simulate values from
the posterior is to draw independent gamma variates $g_k$ with shape
parameters $n_k$ and unit scale parameters and then set $\lambda_k =
g_k/(g_1,\ldots,g_K)$ [\cite{Davison1997}]. There is also a connection
between the posterior distribution for $x$ and bootstrap resampling
[\cite{Rubin1981}].

Since we can assume multiple samples at each covariate value, it is
straightforward to compute the posterior distribution for completely
unstructured priors on the functions $\phi(\cdot)$ and
$\sigma^2(\cdot)$. We introduce vector notation $\phi(\cdot) =
(\phi_1,\ldots,\phi_K)$, $\sigma^2(\cdot) =
(\sigma^2_1,\ldots,\sigma^2_K)$ with
\[
\phi_k =\phi(\xi_k), \qquad\sigma^2_k=\sigma^2(\xi_k),
\]
and independent noninformative prior densities such that
\[
p_{\phi,
\sigma^2}(\phi(\cdot), \sigma^2(\cdot)) = \prod_{k=1}^K p_{\phi_k, \sigma^2_k}(\phi_k, \sigma^2_k)
\]
and
\[ p_{\phi_k,\sigma^2_k} (\phi_k, \sigma^2_k)\propto\sigma^{-2}_k.
\]
It turns
out that for $n_k \geq4$, $\phi_k$ has a posterior $t$-distribution
with easily computable mean and scale parameters. Formulas based on
the data $X$ and $Y$ are given in the Online Supplement
[\cite{Szpiro2010}].

Our main theoretical result is contained in the following theorem,
which is proved in the Online Supplement [\cite{Szpiro2010}]. It states
that, asymptotically, the posterior mean point estimate derived in
equation (\ref{eq:beta.hat}) is the least squares fit to the data $X$
and $Y$, and that the posterior standard deviation from
equation (\ref{eq:sigma.hat}) has the sandwich form. The term
``sandwich'' refers to the algebraic formation in
equation (\ref{eq:sandwich}), where colloquially the $(X^tX)^{-1}$
terms are the ``bread,'' and $(X^t\Sigma X)$ is the ``meat.''
\begin{theorem}\label{th:discrete.pop}
For a discrete covariate space, assume that $y$
conditional on $x$ has bounded first and second moments. The
$m$-dimensional estimate $\bh$ defined by equation (\ref{eq:beta.hat})
takes the asymptotic form
\[
\bh- (X^tX)^{-1}X^tY \rightarrow0,
\]
and assuming there are at least four samples for each covariate
value, the corresponding uncertainty estimate has the asymptotic
sandwich form
%
\begin{equation}\label{eq:sandwich}
\sh_\beta-\D[(X^tX)^{-1}(X^t\Sigma X) (X^tX)^{-1}]^{1/2} =o(n^{-1}),
\end{equation}
where $\Sigma$ is the diagonal matrix defined by
%
\begin{equation}\label{eq:sand}
\Sigma_{ij}=\cases{
\bigl(Y_i - X_i (X^tX)^{-1}X^tY \bigr)^2, &\quad\mbox{if }$i=j$, \cr
0, &\quad \mbox{otherwise.}
}
\end{equation}
The results hold conditionally
almost surely for infinite sequences of observations.
\end{theorem}

\section{Extensions}\label{se:extensions}
\subsection{Continuous covariates}\label{se:cont}
 We consider extending our approach to a continuous
covariate space. The situation is different from discrete covariates
because we cannot expect there to be multiple realizations of each
covariate value in the sampled set. The problem of estimating
$\phi(\cdot)$ and $\sigma^2(\cdot)$ as unconstrained functions is
unidentifiable. However, in applied regression settings it is almost
always reasonable to assume that these are sufficiently regular to be
approximated, using semi-parametric smoothing methods. This is a very
weak assumption compared to assuming linearity and/or homoscedasticity.
We describe a particular choice of spline prior that we implement in
our examples, and leave the general issue of choosing optimal smoothing
priors for future work.

We restrict to scalar $x$ in a model with an intercept, and approximate
$\phi(\cdot)$ and $\log\sigma(\cdot)$ with penalized O'Sullivan
splines using a method based on \cite{Wand2008}, extended to allow for
heteroscedasticity. We pick $Q$ knots spread uniformly over the
potential range of $x$ and set
\begin{eqnarray*}
\phi(\upsilon;u)&=&\alpha_0 + \alpha_1 \upsilon+ \sum_{q=1}^Q u_q B_q(\upsilon), \\
\log\sigma(\upsilon;w)&=&\gamma_0 + \gamma_1 \upsilon+ \sum_{q=1}^Q w_q B_q(\upsilon),
\end{eqnarray*}
where the $B_q(\cdot)$ are B-spline basis
functions defined by the knot locations, with independent priors
$\alpha_i \sim N(0,10^6)$, $\gamma_i \sim N(0,10^6)$. The specification
of priors for $u$ and $w$ involves some transformations and amounts to
the following. Define the matrix $Z$ to incorporate an appropriate
penalty term as in Section 4 of \cite{Wand2008} and let
\begin{eqnarray*}
\phi(X_i;a) & = & \alpha_0 + \alpha_1 X_i + \sum_{q=1}^Q a_q Z_{iq}, \\
\log\sigma(X_i;b)& = &\gamma_0 + \gamma_1 X_i +\sum_{q=1}^Q b_q Z_{iq}
\end{eqnarray*}
with independent priors $a_q \sim N(0,\sigma_a^2)$ and $b_q \sim
N(0,0.1)$ and hyperparameter distributed as $(\sigma_a^2)^{-1} \sim
\textrm{Gamma}(0.1,0.1)$. It is straightforward to simulate from the
posterior distributions using WinBUGS software [\cite{Lunn2000};
\cite{Crainiceanu2005}]. For a prior on the covariate $x$ we use the limiting
case of a Dirichlet process that gives rise to the same posterior
Dirichlet distribution as we had for discrete covariates
[\cite{Gasparini1995}].

An analogous result to Theorem \ref{th:discrete.pop} can be expected to
hold under mild regularity conditions on the true mean and standard
deviation functions $\phi(\cdot)$ and $\sigma^{2}(\cdot)$ in the
data-generating mechanism. We do not state such a result here, but we
provide supporting evidence from a simulation study in
Section~\ref{se:sim}.

\subsection{Fixed design matrix}\label{se:fixed}
Our development up to now explicitly treats $X$ and
$Y$ as being jointly sampled from a random population. The fact that we
obtain an equivalent estimator to the sandwich form suggests that the
sandwich estimator also corresponds to the random $X$ setting. This is
easily seen from equation (\ref{eq:sand}) since the variance estimate
$\Sigma$ in the ``meat'' involves residuals from a linear model and is
bounded away from zero if the data-generating mechanism is nonlinear,
even if the observations $Y$ are deterministic conditional on $X$.

A desirable feature of our approach is that it can easily be modified
to explicitly treat the fixed $X$ scenario. To do this, we simply
replace the random density for $X$ in equation (\ref{eq:beta}) with a
deterministic density corresponding to the actual sampled values
\[
\lambda_{\mathrm{fixed}}(\cdot) = \frac{1}{n} \sum_{i=1}^n\delta_{X_i}(\cdot),
\]
where $\delta_{X_i}$ is the Dirac delta
function with point mass at $X_i$. Then we proceed exactly as in
Section \ref{se:model} to define the quantity of interest
%
\begin{equation}\label{eq:beta.fixed}
\beta_{\mathrm{fixed}} = \argmin\limits_\alpha\int\bigl(\phi(\upsilon) - \upsilon\alpha\bigr)^2 \lambda_{\mathrm{fixed}}(\upsilon)\,d \upsilon.
\end{equation}
The point estimate for fixed $X$ inference is
%
\begin{equation}\label{eq:beta.hat.fixed}
 \hat{\beta}_{\mathrm{fixed}} = E_\pi(\beta_{\mathrm{fixed}} | X,Y),
\end{equation}
and the corresponding measure
of uncertainty is
%
\begin{equation}\label{eq:sigma.hat.fixed}
\hat{\sigma}_{\beta,\mathrm{fixed}} = \D(\C_\pi(\beta_{\mathrm{fixed}}| X,Y) ^{1/2}).
\end{equation}
Notice that the only difference between the
definitions of $\beta$ and $\beta_{\mathrm{fixed}}$ is that in
equation (\ref{eq:beta}) the density $\lambda(\cdot)$ is random while
in equation (\ref{eq:beta.fixed}) the corresponding density is a
deterministic function of the fixed $X$ values.

For the discrete covariate setting we obtain the following result,
which is proved in the Online Supplement [\cite{Szpiro2010}].
\begin{theorem}\label{th:discrete.fixed}
For a discrete covariate space the
$m$-dimensional estimate $\bh_{\mathrm{fixed}}$ defined by
equation (\ref{eq:beta.hat.fixed}) takes the form
\[
\bh_{\mathrm{fixed}} = (X^tX)^{-1}X^tY,
\]
and assuming there are at
least four samples for each covariate value, the corresponding
uncertainty estimate has the sandwich form
\[
\sh_{\beta,\mathrm{fixed}}= \D[(X^tX)^{-1}(X^t\Sigma^\dagger X) (X^tX)^{-1} ]^{1/2},
\]
where $\Sigma^\dagger$ is the
diagonal matrix defined by
\[
\Sigma^\dagger_{ij}=\cases{
\displaystyle\frac{1}{n_k-3}\sum_{l:X_l=\xi_k}(Y_l-\bar{y}_k)^2, &\quad if $i=j$ and $X_i=\xi_k$, \cr 
 0, &\quad  if $i\neq j$
 }
\]
and
 \[
\bar{y}_k=\frac{1}{n_k}\sum_{l:X_l=\xi_k}Y_l.
 \]
\end{theorem}

Since the matrix $\Sigma^\dagger$ in the ``meat'' only includes variation
of $Y$ around its mean, conditional on $X$, this form of the sandwich
estimator appropriately describes sampling variability for fixed $X$,
even if the data-generating mechanism is nonlinear.

\section{Simulations}\label{se:sim}
We consider examples with a single continuous covariate
uniformly distributed in the interval $[-10,10]$, and we evaluate
performance for four true distributions of $y$ given $x$. These are
obtained by taking combinations of the linear response
\[
f_{\mathrm{lin}}(x)= 2+3.5 x
\]
and the nonlinear response
\[ f_{\mathrm{nonlin}}(x) = 2+3.5 x\bigl(1+|\cos(x/2-2)|\bigr)
\]
as well as the equal variance model
$\sigma^2_{\mathrm{equal}} = 5$ and unequal variance model $\sigma^2_{\mathrm{unequal}} =
(5 + x^2/5)$. Example scatterplots of data from each of the four
data-generating models, along with the corresponding mean response
functions, are shown in Figure \ref{fi:examples}.
\begin{table}
\tabcolsep=0pt
\caption{Frequentist properties of estimates for continuous covariate (random $X$)}\label{ta:cont_pop}
\begin{tabular*}{\textwidth}{@{\extracolsep{\fill}}lcccccccc@{}}
\hline
&&&\multicolumn{3}{c}{$\bolds{n=400}$}&\multicolumn{3}{c@{}}{$\bolds{n=800}$}\\
[-7pt]
&&&\multicolumn{3}{c}{\hrulefill}&\multicolumn{3}{c@{}}{\hrulefill}\\
&&&\textbf{Bias}&\textbf{Width}&\textbf{Coverage}&\textbf{Bias}&\textbf{Width}&\textbf{Coverage}\\
\hline
Linear &Equal & Model based & 0.001 & 0.170&0.938&$-$0.001&0.120&0.956\\
&variance& Sandwich&0.001&0.170&0.940&$-$0.001&0.120&0.955\\
&& Bayes robust&0.002&0.177&0.943&\phantom{0,}0.000&0.123&0.959\\
&Unequal &Model based&0.001&0.445&0.859&$-$0.002&0.314&0.863\\
&variance& Sandwich&0.001&0.601&0.948&$-$0.002&0.426&0.957\\
&& Bayes robust& 0.002&0.607&0.955&$-$0.001&0.428&0.956\\ [6pt]
Nonlinear&Equal &Model based&0.001&0.262&0.929&$-$0.001&0.185&0.921\\
&variance& Sandwich&0.001&0.298&0.959&$-$0.001&0.211&0.955\\
&& Bayes robust&0.009&0.289&0.950&\phantom{-0}0.003&0.207&0.950\\
&Unequal &Model based&0.002&0.487&0.859&$-$0.003&0.345&0.865\\
&variance& Sandwich&0.002&0.648&0.959&$-$0.003&0.460&0.952\\
&& Bayes robust&$-$0.030\phantom{0,}&0.657&0.951&$-$0.019&0.460&0.944\\
\hline
\end{tabular*}
\end{table}

For each of the four models we generate $1000$ random realizations of
$X$ and $Y$ with $n=400, 800$. Results are given in
Table \ref{ta:cont_pop} for inference based on random $X$. The
model-based intervals (i.e., standard Bayesian or frequentist linear
regression) fail to give approximate 95\% coverage by being
anti-conservative in all situations except for a linear response with
equal variance. Our Bayesian robust intervals give approximately
correct 95\% coverage for all cases, just like the sandwich intervals.

\begin{table}[b]
\tabcolsep=0pt
\caption{Frequentist properties of estimates for continuous covariate (fixed $X$)}\label{ta:cont_cond}
\begin{tabular*}{\textwidth}{@{\extracolsep{\fill}}lcccccccc@{}}
\hline
&&&\multicolumn{3}{c}{$\bolds{n=400}$}&\multicolumn{3}{c@{}}{$\bolds{n=800}$}\\
[-7pt]
&&&\multicolumn{3}{c}{\hrulefill}&\multicolumn{3}{c@{}}{\hrulefill}\\
&&&\textbf{Bias}&\textbf{Width}&\textbf{Coverage}&\textbf{Bias}&\textbf{Width}&\textbf{Coverage}\\
\hline
Linear &Equal  & Model based & 0.001 & 0.170&0.938&$-$0.001&0.120&0.956\\
&variance& Sandwich&0.001&0.170&0.940&$-$0.001&0.120&0.955\\
&& Bayes robust&0.002&0.173&0.941&\phantom{-0}0.000&0.121&0.954\\
&Unequal &Model based&0.001&0.445&0.859&$-$0.002&0.314&0.863\\
&variance& Sandwich&0.001&0.601&0.948&$-$0.002&0.426&0.957\\
&& Bayes robust& 0.002&0.607&0.951&$-$0.001&0.425&0.953\\ [6pt]
Nonlinear&Equal &Model based&0.001&0.262&0.986&$-$0.001&0.185&0.998\\
&variance& Sandwich&0.001&0.298&0.999&$-$0.001&0.211&1.000\\
&& Bayes robust&0.009&0.187&0.959&\phantom{0,}0.003&0.128&0.963\\
&Unequal &Model based&0.001&0.487&0.893&$-$0.002&0.345&0.888\\
& variance& Sandwich&0.001&0.648&0.961&$-$0.002&0.460&0.968\\
&& Bayes robust&$-$0.030\phantom{0,}&0.629&0.947&$-$0.018&0.437&0.953\\
\hline
\end{tabular*}
\end{table}

We repeat the simulation, treating the observed design matrix $X$ as
fixed. The results are shown in Table \ref{ta:cont_cond}. As
expected, for a linear data-generating mechanism the results are
essentially the same as for random $X$ inference. The model-based
intervals are correct only for the equal variance case, while the
sandwich and Bayesian robust intervals give correct coverage for
unequal variance as well. If the data-generating mechanism is nonlinear
and homoscedastic, then the model-based intervals and sandwich
intervals are conservative since they implicitly account for random
sampling of $X$, while the Bayesian robust intervals give approximately
nominal 95\% coverage. If the data-generating mechanism is both
nonlinear and heteroscedastic, the sandwich intervals are still
slightly conservative when compared to the Bayesian robust intervals,
but both give approximately nominal 95\% coverage. In this situation
the model-based intervals are anti-conservative.

Overall, our Bayesian robust intervals give approximately nominal 95\%
coverage in all situations. The model-based and sandwich-based
intervals can fail by being either conservative or anti-conservative,
depending on details of the data-generating mechanism and the
distinction between random and fixed $X$ sampling.

\section{Health care cost data}\label{se:ex}
 We illustrate our methods using data from the evaluation
of the Washington State Basic Health Plan, as described earlier in
Section~\ref{se:intro} and in more detail by \cite{Diehr1993}. We use
the variable ``cost of outpatient care'' as the outcome and assess its
``linear relationship'' with age. The data are shown in the top panel of
Figure \ref{fi:health}, and O'Sullivan spline fits (see
Section \ref{se:cont}) to the mean and standard deviation as functions
of age are shown in the bottom two panels. The thick red lines are the
posterior means of the Bayesian spline fits, and the thin dashed red
lines are example draws from the posterior distributions.

We can regard the age covariate as either discrete or continuous, and
to illustrate our methodology, we do the analysis both ways. Results
are shown in Table~\ref{ta:health}. The difference in average annual
outpatient health care costs associated with a one year difference in
age is estimated to be 16.1 dollars, with a model-based standard error
of 1.25 dollars. As expected, in light of the heteroscedasticity, the
sandwich form gives a larger standard error estimate of 1.67 dollars.
The uncertainty estimates from our Bayesian robust estimators range
from 1.70 dollars to 1.72 dollars, agreeing very closely with the
sandwich values. The point estimate is nearly identical to the least
squares fit (15.9 dollars) when we model age as continuous in the
Bayesian robust approach; the slight difference is probably due to
approximations involved in fitting the spline model. The Bayesian
robust standard deviations are nearly identical when we model $X$ as
random or fixed, indicating that random variations in average costs
conditional on age contribute more to the uncertainty than does
nonlinearity in the data-generating mechanism.
\begin{table}\tablewidth=320pt
\caption{Linear regression of average annual outpatient health care cost data from
the evaluation of the Washington State Basic Health Plan}\label{ta:health}
\begin{tabular*}{320pt}{@{\extracolsep{\fill}}lcccc@{}}
\hline
&\multicolumn{2}{c}{\textbf{Discrete} $\bolds{X}$}&\multicolumn{2}{c@{}}{\textbf{Continous} $\bolds{X}$}\\ [-7pt]
&\multicolumn{2}{c}{\hrulefill}&\multicolumn{2}{c@{}}{\hrulefill}\\
&$\bolds{\hat{\beta}}$ & $\bolds{\hat{\sigma}_\beta}$ & $\bolds{\hat{\beta}}$ &$\bolds{\hat{\sigma}_\beta}$\\
\hline
Model-based & 16.1 & 1.25 & 16.1 & 1.25\\
Sandwich& 16.1& 1.67& 16.1& 1.67\\
Bayes robust (random $X$)& 16.1 &1.72 &15.9 & 1.71\\
Bayes robust (fixed $X$) &16.1  &1.70 &15.9 & 1.70\\
\hline
\end{tabular*}
\end{table}

\section{Discussion}\label{se:disc}
The main contribution of this paper is a model-robust
Bayesian framework for linear regression that gives uncertainty
estimates equivalent to the sandwich form for random covariate sampling
and with superior sampling properties for a fixed design matrix. In
both situations, our estimates correctly account for heteroscedasticity
and nonlinearity, in the sense of giving asymptotically valid
frequentist sampling properties. The idea is to describe the
data-generating mechanism nonparametrically, and then to define a
functional of the true data-generating mechanism as the quantity of
interest for inference. Once this quantity is defined, we follow common
Bayesian practice and derive a point estimate as its posterior mean and
an uncertainty estimate as its posterior standard deviation. In the
case considered here, the quantity of interest is the least-squares
linear fit to the (potentially nonlinear) mean of the outcome random
variable $y$ conditional on the covariate random variable $x$.

Our conceptual framework is powerful because it provides a general
definition of linear regression in a model-agnostic framework. We can
move seamlessly between classical model-based inference and robust
sandwich-based inference simply by using different priors for
$\phi(\cdot)$ and $\sigma(\cdot)$. If subjective prior information is
available, this can also be included without any modification to the
methodology. Regardless of what information is encoded in the priors,
our target of inference remains the same and has an explicit
interpretation in terms of the trend in the data-generating mechanism.

Our estimation approach transparently distinguishes between the cases
where the observed covariates are regarded as random and where they are
regarded as a fixed design matrix. We obtain good frequentist coverage
properties in both situations, and our estimates are equivalent to the
sandwich form when the covariates are treated as random. For the fixed
design matrix setting, our Bayesian robust intervals can provide
notably better sampling properties than the sandwich estimator in the
situation where the true data-generating mechanism has a mean that is
nonlinear in the covariates. This is true asymptotically, since the
sandwich estimator overestimates the standard errors in this situation
by confusing the part of the residuals that results from nonlinearity
in the data-generating mechanism (which does not vary across samples
and should not contribute to standard error estimates) with the random
component in the residuals (which varies across samples and should be
accounted for in estimating standard errors). This result can be seen
in our simulation examples in Section \ref{se:sim} and by comparing the
expressions for standard errors in Theorems \ref{th:discrete.pop}
and \ref{th:discrete.fixed}. Elsewhere, we have derived similar results
for a fixed design matrix by employing a Bayesian decision theoretic
formalism [\cite{Rice2008}].

In the continuous covariate case, we use splines to approximate the
mean and variance functions for $y$ conditional on $x$. This is
necessary because the mean and variance are not separately identifiable
from a single sample at each covariate value. It can be regarded as a
weakness in our approach, but it also suggests an opportunity to
improve on the small-sample performance of the sandwich estimator by
incorporating additional prior information. Our use of splines will
work in any situation where the true mean and variance are smooth
functions of the covariates. This smoothness is a very reasonable
assumption for applied problems. In fact, the implicit assumption of
the sandwich estimator that the variance function has no structure
whatsoever seems overly permissive. By using properly calibrated
splines or other semi-parametric priors, it should be possible to
improve upon the small-sample performance by borrowing information from
nearby covariate values. This approach appears particularly promising
in the context of generalized estimating equations, where there may be
many samples but too few clusters to accurately estimate a completely
unstructured covariance matrix.

\section*{Acknowledgments}
The authors would like to thank the editor and two anonymous referees
for a number of very helpful suggestions.

\begin{supplement}[id=suppA]
\stitle{Proofs of theorems in ``Model robust regression and a Bayesian
`sandwich' estimator'' (Szpiro, Rice, and Lumley)}
\slink[doi]{10.1214/10-AOAS362SUPP} 
\slink[url]{http://lib.stat.cmu.edu/aoas/362/supplement.pdf}
\sdatatype{.pdf}
\sdescription{We provide proofs of the theorems stated in the paper
``Model robust regression and a Bayesian `sandwich' estimator'' by Adam
A. Szpiro,
Kenneth M. Rice and Thomas Lumley.}
\end{supplement}

\printaddresses

\end{document}